# Effectiveness of flip teaching on engineering students' performance in the physics lab.

Computers & Education 144:103708 (2020)

José A. Gómez-Tejedor[a,1], Ana Vidaurre[a,2], Isabel Tort-Ausina[b,3], José Molina-Mateo[a,4], María-Antonia Serrano[a,5], José M. Meseguer-Dueñas[a,6], Rosa M. Martínez Sala[b,7], Susana Quiles[a,8], Jaime Riera[a,9]

[1]Corresponding author: jogomez@fis.upv.es

[2]e-mail: vidaurre@fis.upv.es

[3]e-mail: isatort@fis.upv.es

[4]e-mail: jmmateo@fis.upv.es

[5]e-mail: mserranj@fis.upv.es

[6]e-mail: jmmesegu@fis.upv.es

[7]e-mail: rmsala@fis.upv.es

[8]e-mail: suquica@fis.upv.es

[9]e-mail: jriera@fis.upv.es

[a]Departamento de Física Aplicada. ETS de Ingeniería del Diseño. Universitat Politècnica de València. Camino de Vera, s/n. 46022 Valencia, Spain.

[b]Departamento de Física Aplicada. ETS de Ingeniería de Edificación. Universitat Politècnica de València. Camino de Vera, s/n. 46022 Valencia, Spain.

**Graphical abstract**

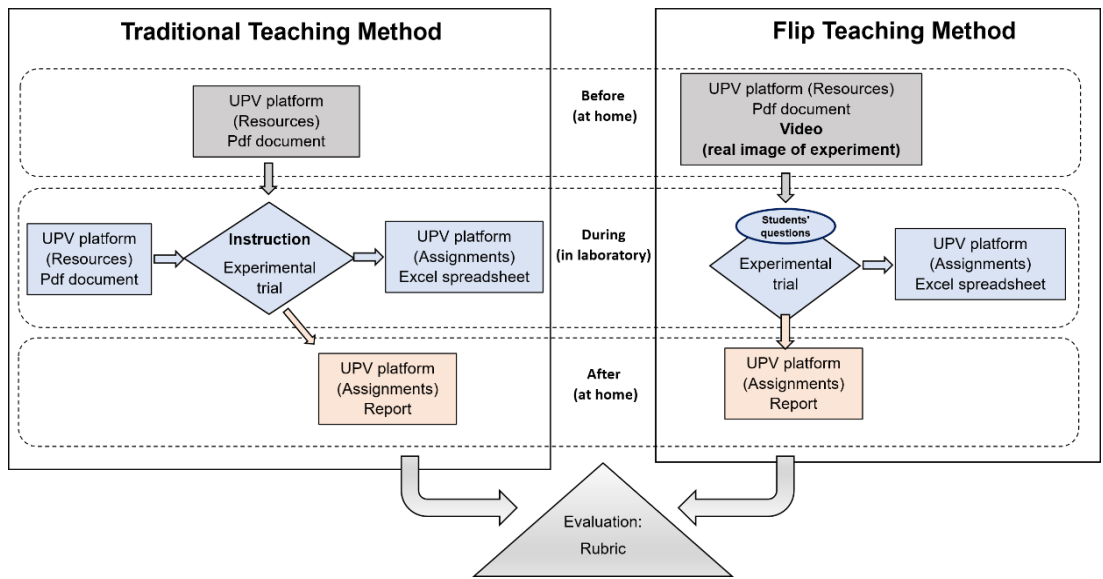

**Highlights:**

- Lab results for Flip and Traditional Methodologies are analysed and compared.
- Flip Teaching in the lab sessions improves students' academic performance.
- Since students spend more time thinking, they write better lab reports.
- Definition of the "analogous students" enables a more detailed statistical study.

**Abstract**

The progressive introduction of the flip teaching (FT) instructional model into higher education has accelerated in recent years. The FT methodology seems to be especially suitable for laboratory practice sessions: before the lab session the students are given documents and videos that explain the theoretical contents and the experimental procedure. When this material is studied in advance, the practice session can be devoted to the discussion, clarification and practical application of the acquired knowledge.

This paper describes the effect of the FT methodology on the students' academic performance when it was applied to the laboratory practice in two subjects, Physics and Electricity, of a technical degree. The laboratory and final grades of these subjects were compared in four consecutive years. The characteristics of all four years were quite similar, except that the traditional teaching method (TM) was used in two, while FT was applied in the other two. The statistical analysis shows that the academic results of the students were better in both subjects under FT than those obtained using TM, and that the difference was statistically significant.

# 1. Introduction

## 1.1. Flipped classroom model

Flip teaching (FT) has become increasingly popular in higher education since it was first described as an "inverted classroom" in 2000 (Lage, Platt, & Treglia, 2000). This method changes the conditions of where and when the students perform their learning tasks. While in the traditional methodology (TM) new concepts are explained by the teacher in class and the student later reflects on the content at home, in the FT model the students prepare the class content alone at home (watching videos or reading texts), while in class they work together and discuss matters as a group (Garrison & Vaughan, 2013; H.-T. Hung, 2015; M.-L. Hung & Chou, 2015). Implementing the FT model involves a change in the instructional design and affects strategies, tools and teaching methods to engage learners in self-directed learning outside the classroom before face-to-face meetings with teachers in the classroom (Teo, Tan, Yan, Teo, & Yeo, 2014). FT is based on active learning models (Bonwell & Eison, 1991; Meyers & Jones, 1993) and is a very general term that means students actually think about what they are doing. It covers a wide range of learning activities, learning strategies and instructional models. In the FT model, the responsibility for learning is transferred from the teacher to the student through participation in interactive activities (Pierce & Fox, 2012). In the FT model the students are engaged in more complex activities while they can obtain teachers feedback in a timely manner (Thai, De Wever, & Valcke, 2017). FT enables teachers to encourage critical thinking in their students, building the capacity for lifelong learning and preparing future graduates for their work-place contexts (Akçayır & Akçayır, 2018; O'Flaherty & Phillips, 2015).

Many different types of flipped models have been developed, each having a different perspective. The flipped classroom thus has many different names and approaches. Some authors consider that terms like "blended learning", "reverse instruction", "inverted classroom", and "24/7 classroom" are interchangeable (Bergmann & Sams, 2012), while others think that FT is a subset of blended learning (Staker & Horn, 2012). In an attempt to identify the mains aspects of the FT schema, the Flipped Learning Network and Pearson's School Achievement Services define the four components that support student's engagement in flipped learning: **F**lexible Environment, **L**earning Culture, **I**ntentional Content and **P**rofessional Educators (F-L-I-P™) (Hamdan, McKnight, McKnight, & Arfstrom, 2013). It has recently been pointed out that these four components are not enough for higher education for a number of reasons, including the fact that the F-L-I-P™ schema does not account for computer learning platforms (Chen, Wang, Kinshuk, & Chen, 2014).

## 1.2. Information and Communications Technology as a basis for flip teaching

Information and Communications Technology (ICT) provides us with the basic tools to create "blended learning" environments, which support flipped classroom model development (Ardid, Gómez-Tejedor, Meseguer-Dueñas, Riera, & Vidaurre, 2015). Computer learning platforms offer a wide range of tools and play an important role in the flip classroom design. Universities have adopted different platforms specifically designed for educational purposes, e.g. that by Moodle and Sakai, which are among the most popular. Rienties et al. (2012) found no correlation between the use of information and communication technologies (ICT) and the teaching approach used in the course design or cognitive content. Zacharia and Olympiou (2011), when comparing the effectiveness of virtual labs *versus* manipulative experimentation, found that both conditions were equally effective in promoting students' understanding of concepts. In other words ICT has spread to all academic levels all over the world. FT takes advantadge of the fact that computer learning platforms can organise the course content into video, tasks and assesment (Ardid et al., 2015). Video recording is one of the most commonly used tools in FT (Pierce & Fox, 2012). The videos proved to be more effective than textbook readings (Jensen, Holt, Sowards, Heath Ogden, & West, 2018) and were preferred by the students (Sahin, Cavlazoglu, & Zeytuncu, 2015). Moreover, Nouri (2016) found that this pre-class material was more positively perceived by the low achievers' group. Others technologies, like augmented reality-based flipped learning or a remote laboratory system, have demonstrated to improve student's performance as well as student's learning motivation (Chang & Hwang, 2018; Tho & Yeung, 2016).

FT is being implemented in a variety of fields, including: physics (Şengel, 2016), maths (Sun, Xie, & Anderman, 2018), chemistry (Baepler, Walker, & Driessen, 2014), chemistry labs (Teo et al., 2014), engineering, social studies and humanities (Kim, Kim, Khera, & Getman, 2014), and economics (Roach, 2014). Most of the papers evaluating student outcomes report improved academic performance over the traditional methodology (TM) (Missildine, Fountain, Summers, & Gosselin, 2013; Pierce & Fox, 2012). Its application to secondary school students was found to induce statistically significant growth in information literacy, competency and critical thinking skills (Kong, 2014, 2015). Although some authors conclude that the flipped classroom does not bring higher learning gains or better attitudes than other methods (Jensen, Kummer, & Godoy, 2015), learning gains are most likely the result of the active-learning style.

In particular, and especially as related to the present paper, FT has been implemented in different physics courses. For instance, Deslauriers, Schelew, and Wieman (2011) found evidence that flipping the classroom can produce significant learning gains by comparing two sections of a large physics class. Şengel (2016) compared the effectiveness of the flipped-classroom approach coupled with problem-based learning and cooperative learning, compared to that of a traditional classroom. The results of this study showed that effectiveness was significantly better in the flipped classroom. In reference (Aşıksoy & Özdamlı, 2016) it was found that an experimental group of students following a flipped classroom model performed better than the students in a group following a traditional lecture format. It can thus be seen that there is general agreement in the literature on the improvement in learning performance in physics courses by applying the FT methodology. Moreover, Vo, Zhu, and Diep (2017) found a higher mean effect size in STEM disciplines as compared to non-STEM disciplines. However, all these studies are centred on theory and problem teaching activities, and there is a lack of studies on the implementation of this methodology in the physics laboratory, which was one of this paper's aims. In this document, we examine the effectiveness of the use of ICT, delivering recorded videos explaining the experiment, that students should prepare before class. This allows more time in the laboratory session to be devoted to the analysis and discussion of the obtained results.

### 1.3. The research goals and questions

In this paper we compare the lab practice results of students in two academic years who followed a TM with those of another group who followed an FT model in the subjects of Physics and Electricity for the Industrial and Automated Electronics Degree at the UPV in the period 2013-2017. The two first academic years (2013-2014 and 2014-2015) followed a TM method, while in the other two (2015-2016 and 2016-2017) the FT method was used. In order to obtain more detailed information on the effect of using the FT method, we matched the students of both methods by defining the *analogous student*. The students were divided into two groups according to their academic results and the normalized gain of analogous students was calculated.

Our hypothesis is that, as we eliminated the teacher's explanation at the beginning of the lab session, if the students can spend more time thinking about what they are doing, asking questions and discussing things with their classmates, they will obtain more detailed knowledge and produce better results. This will also affect the student's motivation and improve their academic results.

In brief, the research questions of this study are:

- Are there any significant differences in learning performance between the FT ICT-based and TM methodology? How does this affect to the different ways of measuring learning performance?
- Do these results affect all students equally, or it depends on their academic level?

## 2. Material and methods
### 2.1. Flip teaching model

This paper describes the methodological changes introduced by FT in the laboratory and analyses their effects on the academic performance of the students in the laboratory. For this, the grades obtained by the students in the years when the FT method was used are compared to the grades obtained in previous years using a TM model. Design plays an important role in both TM and FT and sequencing and progression should be given careful consideration (Jovanovic, Mirriahi, Gašević, Dawson, & Pardo, 2019). In general, it is advisable that the lab-work be directly related (in content and timing) to the rest of the learning activities planned for the subject. This was a key criterion when planning the lab-sessions in TM and later in FT. Neither lab-sessions not teachers were changed when we introduced the FT model.

The lab sessions were organized in the four years as follows: before the session the necessary learning material was made available to students through the Universitat Politècnica de València's online teaching platform, *PoliformaT* ("PoliformaT," 2003), adapted from Sakai ("Sakai Learning Management System," 2018). In this platform students and teachers can share information about their courses and use management tools such as contents resources, assignments, tests & quizzes, etc. In our case, the learning materials consisted of documents explaining the purpose and the procedure of the experience in both cases, and also a video in the FT methodology. The videos were screencasts with the explanation of the work that students must do in the lab session, and the equipment they had to handle. An example of the video introducing the free fall practice can be seen at reference (Molina-Mateo, 2017).

Lab sessions, no matter the methodology used, were based on teamwork. Under continuous teacher supervision, the students worked in groups of six. At the end of each two-hour session the groups uploaded a spreadsheet to the Assignments Section of the *PoliformaT* platform, containing the data acquired and the analytical model used. Each group then had ten days to upload a report. Figure 1 shows a scheme with the main differences between the TM (left) and FT (right) methodologies, based on the teacher explaining the theoretical contents and lab procedure at the beginning of the TM session, while in the FT ones the students had studied the material in advance and the lab session started directly with the experimental work.

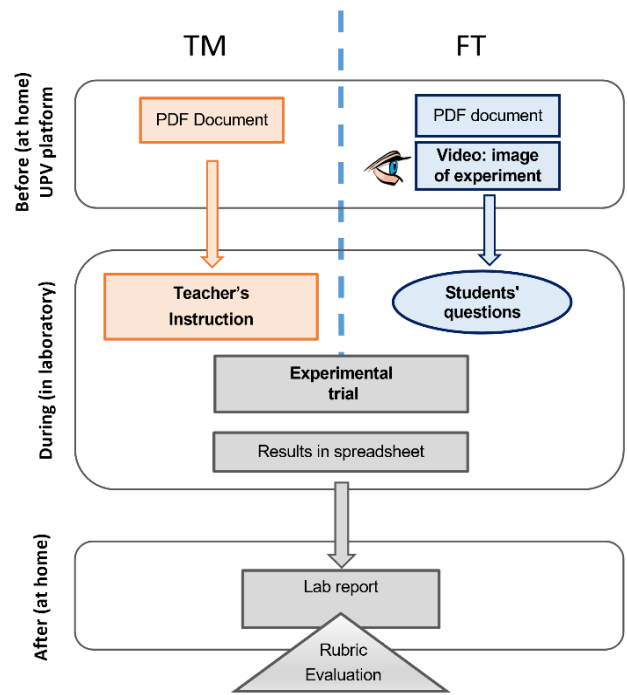

Figure 1. Outline of the differences between TM (left) and FT (right) methodologies. The boxes in the centre represent points common to both methodologies.

There is some discussion in the literature about the convenience of gradually introducing the FT model (Jeffrey Mehring & Leis, 2018; Tomas, Evans, Doyle, & Skamp, 2019). In our case we opted to do this first in the laboratory sessions. We realised that long explanations before the experimental work were a waste of time. FT was introduced into the lab sessions from the first day of class and the teacher's explanations were reduced to answering any doubts about out-of-class learning (reading texts and watching videos). The learning design was based on the "first principles of instruction" (Merrill, 2002) following the scheme of Lo, Lie, and Hew (2018): the videos, as part of the out-of-class learning component, contributes to the activation phase. During the lab sessions the students start measuring under the teacher's supervision and are encouraged to ask questions to be sure they have understood the task (demonstration phase). Later, still in class, they organise, discuss and defend their results (integration phase).

### 2.2. Participants

The sample was formed by 1233 students enrolled from 2013 to 2017 who completed the subjects of Physics and Electricity. The previous grade that gave access to the university has fluctuated between 6 and 7 (out of 10) during all these years. This means that the students' initial level was practically the same in all the years of the study for the TM and FT participants. TM was applied in the first two academic years and FT in the other two. The sample characteristics are summarized in Table 1, also the method used each year and course, the number of enrolled students and the number of students that completed the course.

*Table 1. Number of total enrolled students and number of students that completed the course, academic year and method.*

| Course | Academic year | Method | Number of students enrolled | Number of students following the course |
|--------|---------------|--------|-----------------------------|-----------------------------------------|
| Physics | 2013-2014 | TM | 193 | 159 |
| | 2014-2015 | TM | 172 | 146 |
| | 2015-2016 | FT | 176 | 151 |
| | 2016-2017 | FT | 166 | 148 |
| Electricity | 2013-2014 | TM | 221 | 163 |
| | 2014-2015 | TM | 194 | 152 |
| | 2015-2016 | FT | 191 | 159 |
| | 2016-2017 | FT | 185 | 155 |
| | | | 1498 | 1233 |

The lab report was used to grade the group's lab work (total ten points), which was part (20%) of the final course grade. All the teachers used the same criteria for the years and subjects involved in the study. These criteria (see the rubric at the Supplementary Material) were also available to the students on *PoliformaT* and could be used as a guide to prepare their reports. Moreover, each teacher was lecturing and assessing the same courses in the first years, with TM, and in the latest ones, with FT, that makes the evaluation consistent between the four years analysed.

To analyse the improvement in the students' knowledge from one method to the other, the student sample was segmented according to the final grades obtained.

### 2.3. Data analysis

After anonymization, grades were obtained from the courses in which the authors taught these subjects during the academic years studied. The final grades of the different courses were obtained from the published grades records and the lab grades of each course were collected from the documentation available to the lecturers who taught these subjects.

The laboratory grades were analysed on SPSS software on Windows Version 16 (IBM, Somers, NY, USA) (raw data can be found at the *research data* section). In the first step, a student t-test was applied to the two consecutive years (with the same method) of each course.

Secondly, the possible differences in the average grade of the laboratory between the four groups (FT and TM in both years) were then analysed by a parametric one-way ANOVA test. Its applicability had previously been verified by a Levene test, which showed the homogeneity of the variances of the laboratory grades. The sample was segmented into two groups (high and low performance levels) according to the median of the total grade for the course (final grade total of 10 points). A two-way ANOVA test then analysed the interaction between the effect of the method and the students' level.

### 2.4. Measure of improvement

#### 2.4.1. Cohen´s *d* and average normalized gain.

In order to assess the improvement obtained by the FT methodology, students' learning can be measured by Cohen's *d* that is commonly used in the broader field of educational research. Cohen's *d* is defined as the difference between the average post-test minus pre-test divided by the pooled standard deviation ($s_p$) of the pre- and post-tests (Cohen, 1988):

$$d = \frac{\overline{FT} - \overline{TM}}{s_p} \qquad (1)$$

Where $\overline{TM}$ and $\overline{FT}$ stand for the mean grades (out of 10) for the TM and FT lab grades, respectively.

The most popular metric used in physical education research is the normalized gain (*g*) developed by Hake (1998) and defined as the ratio of the actual average gain to the maximum possible average gain:

$$g = \frac{\overline{FT} - \overline{TM}}{10 - \overline{TM}} \qquad (2)$$

The normalized gain *g* is accepted as a good index to measure the knowledge acquired by the students. Many studies have used this method of measuring students' learning in science and engineering fields (Bao, 2006; Coletta & Phillips, 2005; Marx & Cummings, 2007; Nissen, Talbot, Nasim Thompson, & Van Dusen, 2018; Willoughby & Metz, 2009). Nissen, Talbot, Nasim Thompson, and Van Dusen (2018) compared the gain and effect size metric and concluded that *g* is biased in favour of high pre-test populations and puts groups with lower results at a disadvantage when instructors use *g* to measure instructional efficacy.

Another commonly used method is to calculate the normalized gain for each student, $g_i$, and then average the normalized gains to calculate $\overline{g}$ for the group (Bao, 2006; Von Korff et al., 2016). Using individual gains provides more detailed information than the average value. For instance, it makes possible to find any significant differences between two groups or define the confidence intervals. The group can also be divided into subgroups of students in order to analyse the dependence on different parameters, such as grade level, gender, or others (Kohlmyer et al., 2009; Şengel, 2016). Bao (2006) claims that comparing *g* and the average of $g_i$, $\overline{g}$, can indicate any changes in a group of students, and also that when $\overline{g}$ is lower than *g*, students with low pre-test scores tend to have larger score improvements than those with high pre-test scores, while the opposite happens when $\overline{g}$ is higher than *g*.

#### 2.4.2. The analogous students normalized gain

In our study, as the same students did not follow the two methodologies, we could not calculate the normalized gain for each student. To overcome this difficulty and benefit from a more robust statistical development, we introduced "analogous students" concept between the FT and TM groups (the one that occupies the same order by overall grade) following a two-step process (Figure 2):

1. First, we divide the groups into two levels, low and high, according to the total course grade.
2. To identify the analogous students, the students were ranked according to their lab grades in each of the two subgroups (high or low grades) for each course. Two students were considered analogous when they were in the same position in the ranking, as can be seen in Figure 2.

The introduction of the concept of "analogous students" allowed us to associate confidence intervals with the results of the gain and perform a more robust statistical study.

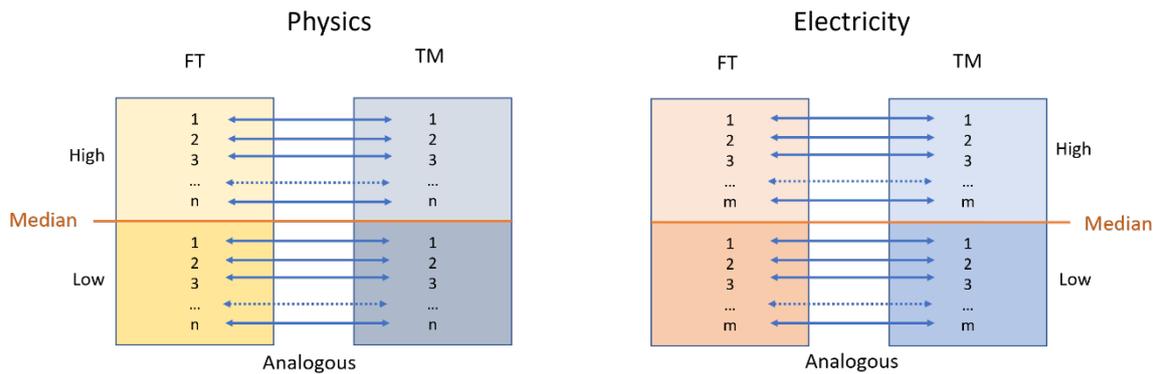

Figure 2. Procedure to define the "analogous students" according to the average laboratory grade.

In cases where the number of students was not the same in the subgroups of the two methodologies, some students were excluded randomly from the subgroup with most members. Table 2 shows the size of the student sample finally used in this analysis.

*Table 2. Size of students sample used to calculate analogous students normalized gain.*

|        | Low grade | High grade | Total |
|--------|-----------|------------|-------|
| PHYS   | 150       | 149        | 299   |
| ELEC   | 157       | 157        | 314   |
| Total  | 307       | 306        | 613   |

Then, for each course, the individual normalized gain of "analogous" students for the FT methodology was calculated according to (Bao, 2006):

$$g_i = \begin{cases} \frac{FT_i - TM_i}{10 - TM_i} & (FT_i > TM_i) \\ \frac{FT_i - TM_i}{TM_i} & (FT_i < TM_i) \end{cases} \qquad (3)$$

Where $FT_i$ and $TM_i$ are the laboratory grades (out of 10) of the corresponding FT and MT analogous students.

From this we established the values of the gain by levels, with their respective confidence interval. The results obtained by this method are compatible with those obtained by traditional methods.

So, the improvement due to FT was defined as an average individual normalized gain $\overline{g}$. The results were compared with the more commonly used Cohen's $d$ and the normalized gain $g$.

## 3. Results and discussion
### 3.1. Application of the FT model

Although some studies have indicated that students were reluctant to accept the FT methodology (Chen et al., 2014; Şengel, 2016), we did not notice any difficulty in adopting this new approach. Although we could not trace how resources had been used in the FT courses studied (at that time there was no tool available on the *PoliformaT* platform), there was evidence for the use of the resources. We found that after the first lab session, the students realized that to be able to carry out the practice correctly they needed to read the pdf documents and watch the videos in advance. In general, the students following the FT model had a more active attitude during the class, asking questions or arguing while defending their point of view.

### 3.2. Statistical analysis of the impact of the methodology

The evolution of the average laboratory grades and the final grade of both courses is shown in Table 3. An increase in the average laboratory grade is observed when the new methodology is introduced. A student t-test was performed on the students that followed the complete course and showed no significant differences between the two TM and FT years, when performed independently for Physics (abbreviated as *Phys*) and Electricity (abbreviated as *Elec*), as can be seen in Table 3.

*Table 3: Average final grade and its standard deviation, average laboratory grade and its standard deviation, t-value and p-value.*

|  | Average final grade | Standard deviation | Average laboratory grade | Standard deviation | *t* | *p-value* |
|---|---|---|---|---|---|---|
| PhysTM (2013/14) | 7.17 | 1.31 | 7.06 | 1.31 |  |  |
| PhysTM (2014/15) | 6.92 | 1.53 | 7.32 | 1.92 | -1.36 | 0.18 |
| PhysFT (2015/16) | 7.20 | 1.28 | 7.50 | 1.58 |  |  |
| PhysFT (2016/17) | 7.44 | 1.19 | 7.47 | 1.21 | 0.18 | 0.86 |
| ElecTM (2013/14) | 5.51 | 1.46 | 6.79 | 1.49 |  |  |
| ElecTM (2014/15) | 5.69 | 1.74 | 7.02 | 1.67 | -1.28 | 0.20 |
| ElecFT (2015/16) | 6.13 | 1.46 | 7.17 | 1.53 |  |  |
| ElecFT (2016/17) | 6.27 | 1.29 | 7.36 | 1.50 | -1.07 | 0.29 |

In order to join students from different academic years using the same learning methodology, we compared the first- and second-year groups and found no significant differences. This fact allowed us to joint these two groups in order to increase the total number of students included in each study group. This means that the sample can be divided into four groups, one for each course and method. The data of the newly defined groups are shown in Table 4.

*Table 4: Number of students (N), average final grade, its standard deviation and median, average laboratory grade and its standard deviation.*

|  | *N* | Average final grade | Standard deviation | Median final grade | Average laboratory grade | Standard deviation |
|---|---|---|---|---|---|---|
| PhysTM | 305 | 7.05 | 1.42 | 7.10 | 7.18 | 1.63 |
| PhysFT | 299 | 7.32 | 1.24 | 7.30 | 7.49 | 1.41 |
| ElecTM | 315 | 5.60 | 1.60 | 5.60 | 6.90 | 1.59 |
| ElecFT | 314 | 6.20 | 1.38 | 6.00 | 7.27 | 1.51 |
| Total | 1233 | 6.53 | 1.57 |  | 7.21 | 1.55 |

We then tried to find significant differences between the groups. As Levene's test showed the homogeneity of variances of the laboratory grades of the 4 groups (Levene's statistic was 2.46), a parametric ANOVA test was applied and revealed significant differences between the groups ($F_{(3, 1229)}=7.57$ and $p<0.001$).

To delve into these differences, orthogonal contrasts were performed. The null hypotheses tested were:

1) The Physics laboratory grade of using TM is equal to the grade using FT (Phys TM = Phys FT).

2) The Electricity laboratory grade of using TM is equal to the grade using FT (Elec TM = Elec FT).

3) The laboratory grade of both courses together using TM is equal to the grade using FT (Grade TM = Grade FT).

The obtained results for these hypotheses made the contribution of the methodology evident:

- There were significant differences in the course of Physics (Phys FT > Phys TM; $t=2.44$; $p<0.05$).

- There were significant differences in the course of Electricity (Elec FT > Elec TM; $t=-2.94$; $p<0.01$).

- There were significant differences between both methodologies (FT > TM; $t=-3.81$; $p<0.001$).

This analysis showed the positive impact of the FT methodology on the academic results in both courses as the FT results are significantly better than those of TM.

This latter result is in line with previous studies in different fields, as for instance in nursing education (Missildine et al., 2013) and pharmaceutical students (Pierce & Fox, 2012). The same results have been found in the physics field (Aşıksoy & Özdamlı, 2016; Deslauriers et al., 2011;

Şengel, 2016), although unlike these studies, our work is focused on the application of this methodology in the physics lab.

### 3.3. Interaction effect between the methodology and the subgroups segmented by level

The next step was to determine whether this effect was homogeneous for students with different academic results. For this the students were classified according to their final grades, using the median shown in Table 3. The sample was divided into eight groups according to the subject (Physics or Electricity), the methodology (TM or FT), and their final grade. Table 5 shows the characteristics of theses eight groups, including the average lab grade.

*Table 5: Number of students (N), average laboratory grade (ALG) of the eight groups of students.*

|  | PHYS TM | | PHYS FT | | ELEC TM | | ELEC FT | | Total |
|---|---|---|---|---|---|---|---|---|---|
|  | N | ALG | N | ALG | N | ALG | N | ALG | N |
| Low grade | 153 | 6.46 | 150 | 6.98 | 158 | 6.31 | 157 | 6.59 | 618 |
| High grade | 152 | 7.91 | 149 | 8.00 | 157 | 7.50 | 157 | 7.94 | 615 |
| Total | 305 | | 299 | | 315 | | 314 | | 1233 |

The improvement of the laboratory grades (see Table 5) with FT is also shown in Figure 3, for both subjects. It can be observed that the TM results are always lower than those of FT for the same group. It can also be seen that the groups with the highest final grades also obtained the highest grade in the laboratory, with no interaction between the variables.

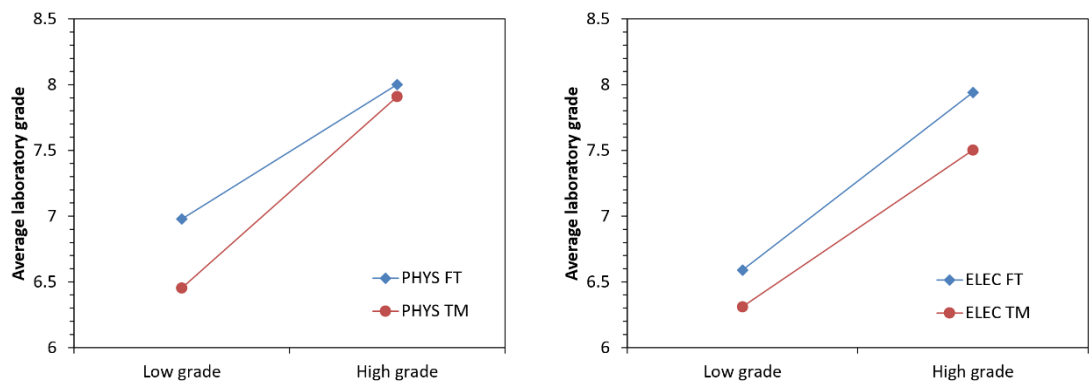

Figure 3: Average laboratory grade for the two methodologies in the two levels, in Physics and Electricity

To check whether the differences observed were statistically significant, a two-way ANOVA analysis was performed (see Table 6). The average Physics laboratory grade, that was higher than the grade in Electricity (see Table 3), shows significant differences between the low- and high-grade subgroups, and also between both methods, while the interaction effect between both variables has no significant influence ($F_{(1, 603)}=3.55$, $p=0.06$).

Similar results were found for Electricity: the grade subgroup and method have a significant influence on the average laboratory grade, while the interaction between them has no significance ($F_{(1,628)}=0.50$, $p=0.48$).

*Table 6: Results of the ANOVA analysis, where SG stands for the subgroup grade, M stands for methodology, SS stands for sum of squares, DF stands for the degrees of freedom.*

|  | Source | SS | DF | F | p |
|---|---|---|---|---|---|
| Physics | SG | 230.91 | 1 | 119.07 | <0.001 |
|  | M | 14.81 | 1 | 7.26 | 0.007 |
|  | (SG)*(M) | 6.88 | 1 | 3.55 | 0.060 |
| Electricity | SG | 254.01 | 1 | 126.90 | <0.001 |
|  | M | 20.29 | 1 | 10.12 | 0.002 |
|  | (SG)*(M) | 1.01 | 1 | 0.50 | 0.480 |

The results obtained allow us to conclude that the method has a clear influence on the students' laboratory grades. In addition, the laboratory grade depends on the student's subgroup grade (high or low). However, there is no interaction between the two factors (method and subgroup grade).

### 3.4. Estimation of the improvement of the academic performance by groups of students according to their grade

#### 3.4.1. Cohen´s *d* and average normalized gain.

The results of the Cohen's *d* calculations are shown in Table 7. The values obtained confirm that FT method improved the students' lab performance and therefore in the course, with a medium effect of between 0.3-0.5, in the low-grade Physics group (0.321) and in the high-grade Electricity group (0.394), with a lesser effect on the other two groups.

Table 7 shows the result obtained for *g*. It is common to categorize the data obtained in three zones in the following way: low, if it is less than 0.3; medium, if it is between 0.3-0.7 and high if it is greater than 0.7 (Hake, 1998). These values also show that although the net gain is low, all the students are positively affected by the FT method.

In accordance with the Cohens' *d* results, the highest average normalized gain is found in the low-grade group for Physics (14.7%), and in the high-grade group for Electricity (17.6%).

*Table 7. Average and statistical parameter of the Cohen´s d and normalized gain g for each grade level and course.*

| Course-Grade level | Cohen´s *d* | *g* |
|---|---|---|
| PHYS Low | 0.321 | 0.147 |
| PHYS High | 0.081 | 0.043 |
| ELEC Low | 0.144 | 0.076 |
| ELEC High | 0.394 | 0.176 |

### 3.4.2. Average of the analogous students normalized gain.

The results of $g_i$ calculated according to Eq. (3) are shown in Figure 4. The $g_i$ values are mostly between 0 and 0.2, except for the highest values of the TM scores. In this region, the results are scattered and are different for Physics and Electricity. In the latter subject the gain of the group with the highest TM scores is higher than that of the lowest TM scores. In Physics the gain does not show a clear dependence on the TM scores.

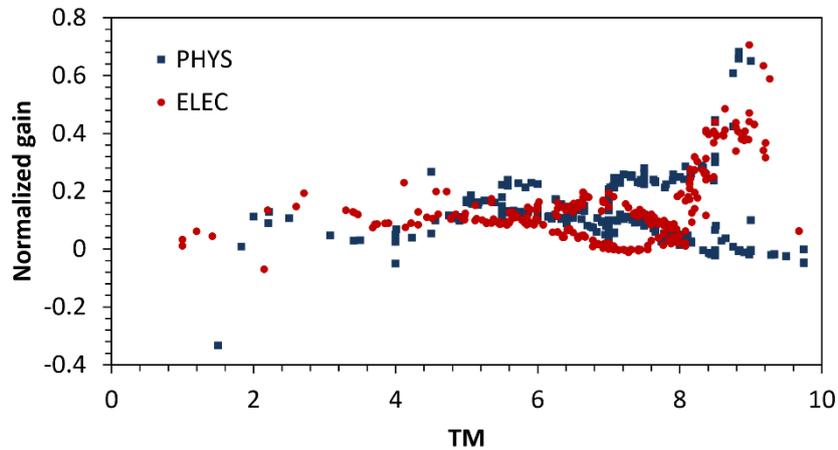

Figure 4: Individual normalized gain for FT methodology for both courses *vs* the laboratory grade in the TM

Next, we calculated the average of the individual normalized gain, $\overline{g}$ for each grade subgroup with its confidence interval. The results are shown in Table 8, for both subjects. They show that there are significant differences between the low- and high-level groups; with a higher gain for the low-grade group than for the high-level group. However, in Electricity the high-level group obtains a greater gain. These results are in complete agreement with those previously obtained for the size effect, as measured by Cohen's *d* or the normalized gain *g* (Table 7).

*Table 8. Average normalized gain ($\overline{g}$), standard deviation (SD),* Std. Error Mean, Confidence interval (CI), student t-test p-values.

| Course-Grade level | Average normalized gain ($\overline{g}$) | SD | Std. Error Mean | CI | student t-test p-value |
|---|---|---|---|---|---|
| PHYS Low | 0.17 | 0.12 | 0.010 | (0.15, 0.19) | <0.001 |
| PHYS High | 0.06 | 0.12 | 0.005 | (0.04, 0.07) | |
| ELEC Low | 0.09 | 0.11 | 0.009 | (0.07, 0.11) | <0.001 |
| ELEC High | 0.20 | 0.12 | 0.009 | (0.18, 0.22) | |

These results are in agreement with those previously obtained: the highest gain is obtained in the low-grade group in Physics (17%), and in the high-grade group in Electricity (20%), which confirms that although the gain is low the students are positively affected by the FT method.

One of the possible explanations is that the videos used to show the experimental procedure, together with the longer time spent in discussion and experimental work in the lab, significantly improved all the levels and subjects, especially in the lowest level of physics students.

On the other hand, while the Electricity laboratory has a higher difficulty and the Physics lab has a lower level of difficulty, physics performs like a normal subject, with the results already described in the bibliography, in which the lowest-level group has the highest gain, since the high-level saturates (Nouri, 2016). As the high level does not saturate in Electricity, both levels improve when the methodology is changed, because the higher-level already has scope for improvement, which does not happen in the physics group.

## 4. Conclusions

This paper studies the influence of introducing the flipped-classroom model into ICT-based laboratory sessions in the subjects of Physics and Electricity in the Industrial and Automated Electronics Degree course at the Universitat Politècnica de València. Data, covering the period 2013-2017 following the TM method in the two first years (2013-2014 and 2014-2015) and the FT method in the other two (2015-2016 and 2016-2017), has been analysed.

The main changes brought about by the FT methodology were that videos were added to the materials available to the students before the lab session, and that the teachers' explanation in the lab was replaced by the students' preparation at home (using the available material). The students therefore had more time in the lab session to organise, discuss and defend their results.

### 4.1. Educational implications

The results show that there are no significant differences between the academic results of consecutive academic years in a subject when the same teaching method is used. However, when the sample is divided into four groups according to the course and the teaching method, there are statistically significant differences between the academic results, as higher laboratory grades were always achieved by the FT method. The Cohens'd and normalized gain results show that all the students are positively affected by the FT method. In the FT methodology, the time devoted to teacher instruction was moved to start working; leaving extra time for more productive activities (demonstration and activation phases). Therefore, FT methodology application to laboratory practices has resulted in a significant improvement in academic results.

Once the influence of the methodology had been studied, the next step was to analyse whether this effect was homogeneous for students with different academic results. After the segmentation analysis of each course, we concluded that, regardless of the statistical method used, the FT method improves the academic results for both levels. The gain is higher for the high-level Electricity group and in the low-level Physics group. Our study has thus shown that academic improvement does not affect all the students in the same way, and it affects to a greater extent the students of the lower level in Physics and students of high level in Electricity. We should point out that we have introduced an innovative method of finding analogous students to obtain the average individual normalized gain. By doing this we have obtained more detailed information on the gain than from the normalized gain alone. We propose this methodology to study similar cases in which methodological changes on the same students cannot be experimented.

As a general conclusion, FT works very well for lab sessions in technical degree courses, confirming the initial hypothesis of the teachers involved. Students work better with the FT method in the laboratory and get better academic results. It also highlights the role of the teacher in enhancing students' autonomous learning. For a progressive application of FT, we recommend starting the experience with the laboratory practices.

### 4.2. Limitations and suggestions for future research

Our study was applied to the subjects of Physics and Electricity for the Industrial and Automated Electronics Degree and indicates that the improvement does not affect both subjects in the same way. The reason for this discrepancy can be interpreted as the differences in the difficulty level of both subjects. In the more difficult subject of Electricity, the high level does not saturate, so that both levels improve when changing the methodology, since the higher-level already has scope for improvement, which does not happen with the Physics group. This shows up the study's limitations and encourages us to continue research on this point. However, caution should be used in extrapolating the results to non-stem disciplines.

The good overall results obtained by the FT indirectly show the effectiveness of using computers in education and encourages research into new ICT-based tools to support the flipped classroom model in an autonomous learning environment.

### Acknowledgements

We would like to thank the Instituto de Ciencias de la Educación (ICE) in the Universitat Politècnica de València for their help, through the Innovation and Educational Quality Program and for supporting the team Innovación en Metodologías Activas para el Aprendizaje de la Física (e-MACAFI).

### Funding

This work was supported by theUniversitat Politècnica de València [Project PIME/2018/B25 Convocatoria de Proyectos de Innovación y Convergencia de la UPV].